# Ferromagnetic semiconductor (In,Ga,Mn)As with Curie temperature above 100 K

**T. Slupinski,[a] H. Munekata,[b] and A. Oiwa**

***Kanagawa Academy of Science and Technology***

***3-2-1 Sakado, Takatsu, Kawasaki 213-0012, Japan***

**Abstract**

We have grown $(In_yGa_{1-y})_{1-x}Mn_xAs$ ferromagnetic semiconductor layers with Mn composition $x$ up to 0.13 on InP substrates by molecular beam epitaxy. Near the lattice-matched composition, i.e., $y$ ~ 0.53, the Curie temperature increases linearly with the ferromagnetically effective Mn composition $x_{eff}$, following the empirical equation $T_C = 1300 \times x_{eff}$. We obtained Curie temperatures above 100 K when $x$ is relatively high ($x > 0.1; x_{eff} \geq 0.08$) and the hole concentration is in the order of $10^{19}$ cm$^{-3}$.

PACS number: 61.72.Vv, 75.50.Pp, 72.80.Ey

---

[a] On leave from Institute of Experimental Physics, Warsaw University, Warsaw, Poland.

[b] Permanent address: Imaging Science and Engineering Laboratory, Tokyo Institute of Technology, Yokohama, Japan.

III-V-based diluted magnetic semiconductors (III-V DMS) (also known as III-V magnetic alloy semiconductors) have attracted much attention as candidate materials for spin-related electronic/optical/magnetic devices. Up to now, (In,Mn)As[1,2,3] and (Ga,Mn)As[4] are the two well-studied semiconductors. The common feature of these two semiconductors is the existence of ferromagnetic order induced by holes supplied by the Mn acceptors.[5] For a given Mn composition, (Ga,Mn)As exhibits a higher Curie temperature than (In,Mn)As, whereas (In,Mn)As can accommodate a larger amount of Mn atoms into the host crystal lattice.[1,5,6] Since it is known that the ferromagnetic order in III-V DMS depends not only on the Mn content and hole concentration but also on the less understood structural state of a metastable alloy determined by sample growth conditions, it should be interesting and important to study experimentally how the ferromagnetism is influenced by alloying these two semiconductors (Ga,Mn)As and (In,Mn)As.

In this Letter, we study $(In_yGa_{1-y})_{1-x}Mn_xAs$ epilayers grown on InP, in the nearly-lattice-matched region, i.e., $y \sim 0.53$. We demonstrate that, in contrast to (Ga,Mn)As,[4] the Curie temperature $T_C$ does not saturate with increasing Mn content $x$ above $x \approx 0.05$, but rather continues to increase even in a relatively high $x$ region ($x > 0.1$). In particular, we have obtained Curie temperatures of $T_C = 100-120$ K for $x = 0.13$, which is comparable to or higher than the maximum $T_C$ value achieved to date in (Ga,Mn)As ($T_C = 100-110$ K for $x \approx 0.05$[4,7]).

We grew $(In_yGa_{1-y})_{1-x}Mn_xAs$ layer with In content $y = 0.53$ on InP(100) substrates by molecular beam epitaxy (MBE). The range of Mn content was $x = 0.03-0.13$. First, a 100-nm thick $In_{0.53}Ga_{0.47}As$ buffer layer was grown at the substrate temperature of $T_s$ = 460 °C. This was followed by the low-temperature epitaxial growth (LT-growth) of a 50-nm thick $(In_{0.53}Ga_{0.47})_{1-x}Mn_xAs$ layer at the substrate temperature of $T_s = 200-260$ °C. High and low-temperature growth conditions were based on the established procedures.[8,9] Similar to the growth of (Ga,Mn)As,[4] we found that the optimum $T_s$ for LT-growth of ferromagnetic (In,Ga,Mn)As tends to decrease with increasing $x$. For this reason, $T_s$ during the LT-growth was controlled by an infrared pyrometer (Ircon Inc., Modline 3, wavelength 2.0-2.6 micrometer). The ratio of V/III beam-equivalent pressures during LT-growth was $As_4$/(In+Ga) = 7-10. The growth rate was relatively high, about 1 monolayer/sec. The $x$ and $y$ values were controlled by the Ga, In and Mn arrival rates (atoms/cm$^2$sec) calibrated based on the growth rates of GaAs, InAs and MnAs layers. For comparison purpose, we also prepared $(In_yGa_{1-y})_{1-x}Mn_xAs$ layers with $y < 0.2$ (close to (Ga,Mn)As) and $y > 0.9$ (close to (In,Mn)As) on GaAs(001) substrates with GaAs or GaSb buffers, respectively, under the growth conditions described in separate papers.[10]

During the epitaxial growth of ferromagnetic $(In_{0.53}Ga_{0.47})_{1-x}Mn_xAs$ layers, a diffused (1 x 1) or (1 x 2) streaky pattern was observed by the reflection high energy electron diffraction (RHEED). An example is shown in the inset of Fig. 1 for $x = 0.13$. The RHEED intensity oscillations were observed, as shown in Fig. 1, for specular reflection during the growth of the $In_{0.53}Ga_{0.47}As$ buffer layer at $T_s = 460$ °C and the $(In_{0.53}Ga_{0.47})_{0.87}Mn_{0.13}As$ layer at $T_s = 200$ °C. The Mn composition estimated from the difference in the oscillation periods between the growths with and without a Mn beam flux is $x = 0.15$ which is close to the value based on the arrival rates.

The $(In_{0.53}Ga_{0.47})_{1-x}Mn_xAs/In_{0.53}Ga_{0.47}As/InP(001)$ samples grown under the above described conditions showed ferromagnetic behavior in magnetization measurements. In addition, ferromagnetism manifested itself in Hall effect measurements, due to the strong correlation between magnetic and electrical properties in ferromagnetic semiconductors. Figure 2 shows Hall resistance curves at various temperatures for a $(In_{0.53}Ga_{0.47})_{0.87}Mn_{0.13}As$ sample having the Curie temperature above 100 K. Similar to (In ,Mn)As and (Ga,Mn)As, the anomalous (or extraordinary) term $R_sM$ is the dominant component in the measured Hall resistivity[1,4] $\rho_H = R_H d = R_o B + R_s M$ , where $R_o$ and $R_s$ are the ordinary and anomalous Hall coefficients, respectively, $R_H$ is the Hall resistance, $d$ is the magnetic layer thickness, and $B$ and $M$ are the magnetic flux density and magnetization perpendicular to the sample surface. Magnetization curves characteristic of a sample with ferromagnetic order can be clearly recognized in Fig. 2. The data show that the remanent magnetization can be observed up to at least 100 K. At about 120 K, the remanent component ceases to exist, indicating the disappearance of ferromagnetic order. One may notice that even above the Curie temperature the Hall resistance curves still show strong nonlinear behavior, and e.g., the sign of the Hall coefficient changes from positive to negative at around 200 K. This complicated transport behavior is probably due to the spin-related scattering that takes place even at room temperature for such a high Mn composition.

We also carried out magnetization measurements with a dc-SQUID magnetometer. High-quality ferromagnetic samples exhibited well-defined squared hysteresis loops with coercive field $H_c$ = 50–100 Oe. The magnetization easy axis was in the plane of the epilayer, presumably due to the biaxial compressive strain in the layer.[6]

The temperature dependence of the magnetization obtained under a weak field ($H$ = 50 Oe) is shown in the upper inset to Fig. 2. The Curie temperature $T_C$ about 110 K is seen, consistently with the transport data. We have also tried to estimate the Curie temperature by the Arrott plot method, using the magnetization determined through the transport data. Assuming that for a relatively high Mn content the side-jump scattering predominates,[11] the anomalous Hall term $R_sM$ can be re-written as $c \cdot {R_{sheet}}^2 \cdot M$ ($c$ constant, $R_{sheet}$ sheet resistance (ohm/ )). An Arrott plot, namely, $({R_H}^a / {R_{sheet}}^2)^2$ vs. ${R_{sheet}}^2 \cdot B / {R_H}^a$, where ${R_H}^a = R_H - R_o B / d$ , is shown in the lower inset to Fig. 2. The ordinary Hall term, $R_o = 1/pe$, was determined by the method described in the next paragraph. Because the line for 120 K consists of both convex and concave parts, it seems ambiguous and we are not able to determine precisely the Curie temperature. However, the line for 100 K shows that $T_C$ value is higher than 100 K.

We now turn to the calculation of the hole concentration $p$ through a fitting of the Hall resistance $R_H$ at 4 K (Fig. 2), using the magnetoresistance $R_{sheet}$ and SQUID magnetization data $M$ up to 7 Tesla. First, the anomalous Hall coefficient $R_s(0)$ has been determined by the small field ($B \approx 0$) values of the magnetization and the Hall resistance at 4 K, where the ordinary Hall term $R_o B$ is much smaller than the anomalous one, $R_s M$. We divided the Hall resistivity $R_H d$ at $B \sim 0.1$ Tesla by the low-field saturation magnetization $M$ at $B \sim 0.1$ Tesla [12]. Then, we calculated the field dependence $R_s(B) = R_s(0) \cdot R_{sheet}{}^2(B) / R_{sheet}{}^2(0)$, scaling up the low-field anomalous Hall coefficient $R_s(0)$ with the square of resistivity to a high magnetic field region, assuming the dominance of the side-jump scattering mechanism mentioned above. Here, $R_{sheet}$ exhibited negative magnetoresistance, which manifests itself in both $R_s(B)$ and $R_H$. Finally, the ordinary component $R_o B/d$ was calculated by subtracting the anomalous Hall term $R_s(B)M/d$ from the measured Hall resistance $R_H$. The hole concentration calculated in this way is $p = 7 \pm 3 \times 10^{19}$ cm$^{-3}$. This value is lower than the one measured for ferromagnetic $p$-(Ga,Mn)As with $T_C = 110$ K, namely $p = 3.5 \times 10^{20}$ cm$^{-3}$ for $x = 0.05$.[5]

It is interesting to notice that the obtained hole concentration ($p = 7 \pm 3 \times 10^{19}$ cm$^{-3}$) is significantly lower than the Mn atom concentration, which is $2.6 \times 10^{21}$ cm$^{-3}$ for $x = 0.13$. This discrepancy cannot be attributed to an electrical compensation by As-related donors since the amount of excess As has been reported[9] to be less than 0.5 at.% ($\sim 1 \times 10^{20}$ cm$^{-3}$) in the $In_{0.53}Ga_{0.47}As$ layers grown by MBE at the substrate temperatures similar or lower than used in the present work. We briefly mention another possibility to describe this discrepancy. Recently, a new idea has been proposed to explain the saturated carrier concentration in the ultra-high doping regime in several semiconductors in terms of the localization of carriers caused by the chemical ordering of impurity atoms in close lattice sites.[13,14,15,16] If this picture also holds for $(In_{0.53}Ga_{0.47})_{1-x}Mn_xAs$ with high Mn contents, it would mean that certain chemical ordering of Mn atoms might result in the reduced hole concentration, even before the occurrence of second structural phases. Such particular chemical ordering, which should probably depend on the sample growth conditions of a metastable alloy, might also affect the ferromagnetic properties. One of the fingerprints that is consistent with this inference is that the *effective* Mn content $x_{eff}$ deduced[17] from the low-field saturation magnetization depends on the growth temperature. Also, our finding that ferromagnetic quaternary samples with high $x_{eff}$ values could be obtained for a limited range of growth temperatures, even narrower than that known for (Ga,Mn)As,[4,18] might be consistent with this picture.

Finally, we discuss some trends in the composition dependence of the ferromagnetic properties of $(In_yGa_{1-y})_{1-x}Mn_xAs$. The SQUID data for samples with different $x$ indicated that the strength of ferromagnetic interaction increased with the ferromagnetically effective Mn content $x_{eff}$. In Figs. 3 (a) and (b), we summarize, in terms of $x_{eff}$, the ferromagnetic characteristics of $(In_yGa_{1-y})_{1-x}Mn_xAs$ alloys near the $y = 0.53$ composition, and also close to (Ga,Mn)As ($y < 0.2$) and (In,Mn)As ($y > 0.9$). We see in Fig. 3 (a) that the $T_C$ of the quaternary alloy in the $y = 0.53$ region appears to be equal or higher than those obtained from both (In,Mn)As and (Ga,Mn)As. Notice that the Curie temperature increases almost linearly with $x_{eff}$, i.e., $T_C = 1300 \times x_{eff}$ and $T_C = 2000 \times x_{eff}$ for $y = 0.53$ and (Ga,Mn)As,[4] respectively. The smaller linear constant for quaternary alloys ($y$ = 0.53) indicates that ferromagnetic Mn-Mn interaction is weaker than that for (Ga,Mn)As. On the other hand, the successful accommodation of higher Mn composition ($x$ = 0.13) achieved in this work for $(In_{0.53}Ga_{0.47})_{1-x}Mn_xAs$ may suggest some positive role of InAs in stabilizing the metastable structure of Mn-containing III-V DMS.

Authors gratefully acknowledge J. Yoshino at Tokyo Institute of Technology (TIT) for the collaboration in high-field magnetotransport experiments, S. Yanagi at TIT for his assistance in molecular beam epitaxy, and J. Kono at Rice University for critical reading of the manuscript. A.O. would like to acknowledge partial support by the Minister of Science, Education, Sport and Culture (Grant-in-Aid No. 12750021).

Figure captions

Fig. 1.

RHEED intensity oscillations observed during the growth of an $(In_{0.53}Ga_{0.47})_{0.87}Mn_{0.13}As$ layer at $T_s$ = 200 °C on an $In_{0.53}Ga_{0.47}As$ buffer layer, and during growth of an $In_{0.53}Ga_{0.47}As$ layer at $T_s$ = 460 °C on an InP(001) substrate. Inset shows a RHEED pattern observed along the [110] azimuth during the growth of a quaternary layer with $x$ = 0.13.

Fig. 2.

Hall resistance curves up to 10 Tesla for $(In_{0.53}Ga_{0.47})_{0.87}Mn_{0.13}As$ layer at temperatures ranging from 4 to 300 K. The magnetic field was applied perpendicular to the sample plane. Insets shows an Arrott plot obtained from magneto-transport data (lower right), and temperature dependence of magnetization measured by SQUID under the application of a weak magnetic field of $B$ = 50 Oe.

Fig. 3.

(a) A plot of Curie temperature vs. In composition $y$ for different effective Mn compositions $x_{eff}$, (b) a plot of Curie temperature as a function of $x_{eff}$ (defined in the text). Solid circles are for $(In_{0.53}Ga_{0.47})_{1-x}Mn_xAs$, whereas open circles for $Ga_{1-x}Mn_xAs$. Dashed lines indicate the relation $T_C = 1300 \times x_{eff}$ for $(In_{0.53}Ga_{0.47})_{1-x}Mn_xAs$ and $T_C = 2000 \times x_{eff}$ for $Ga_{1-x}Mn_xAs$.

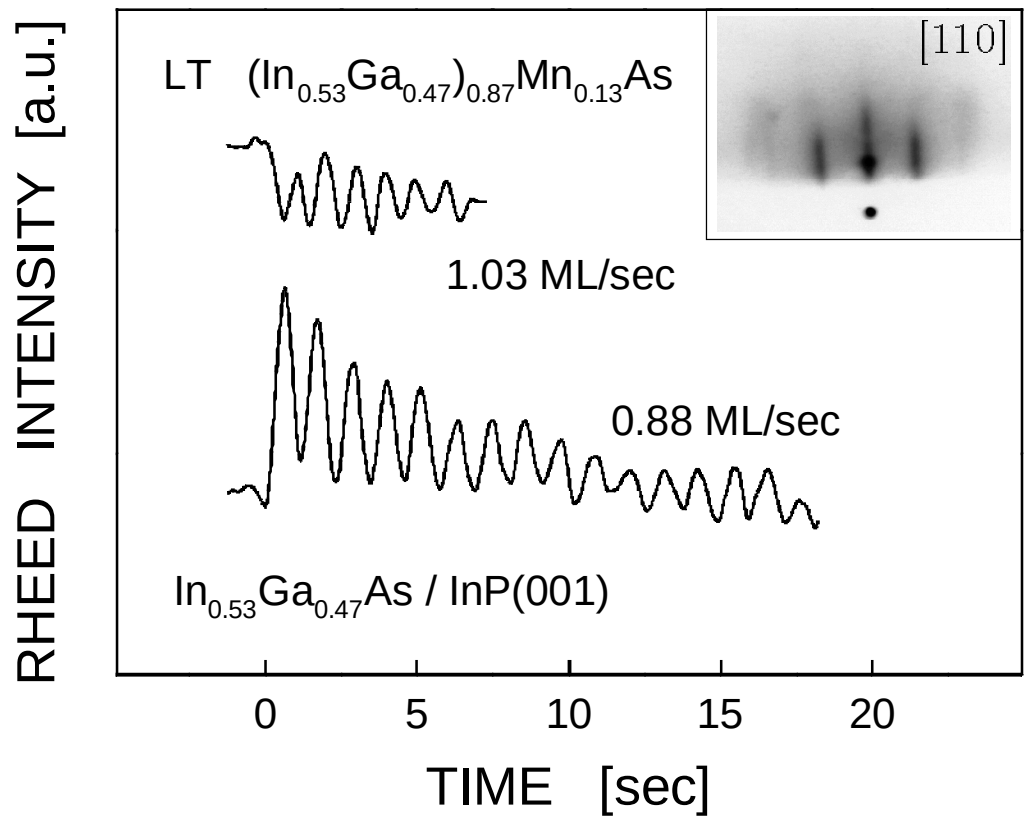

Fig.1.

T. Slupinski et al., APL

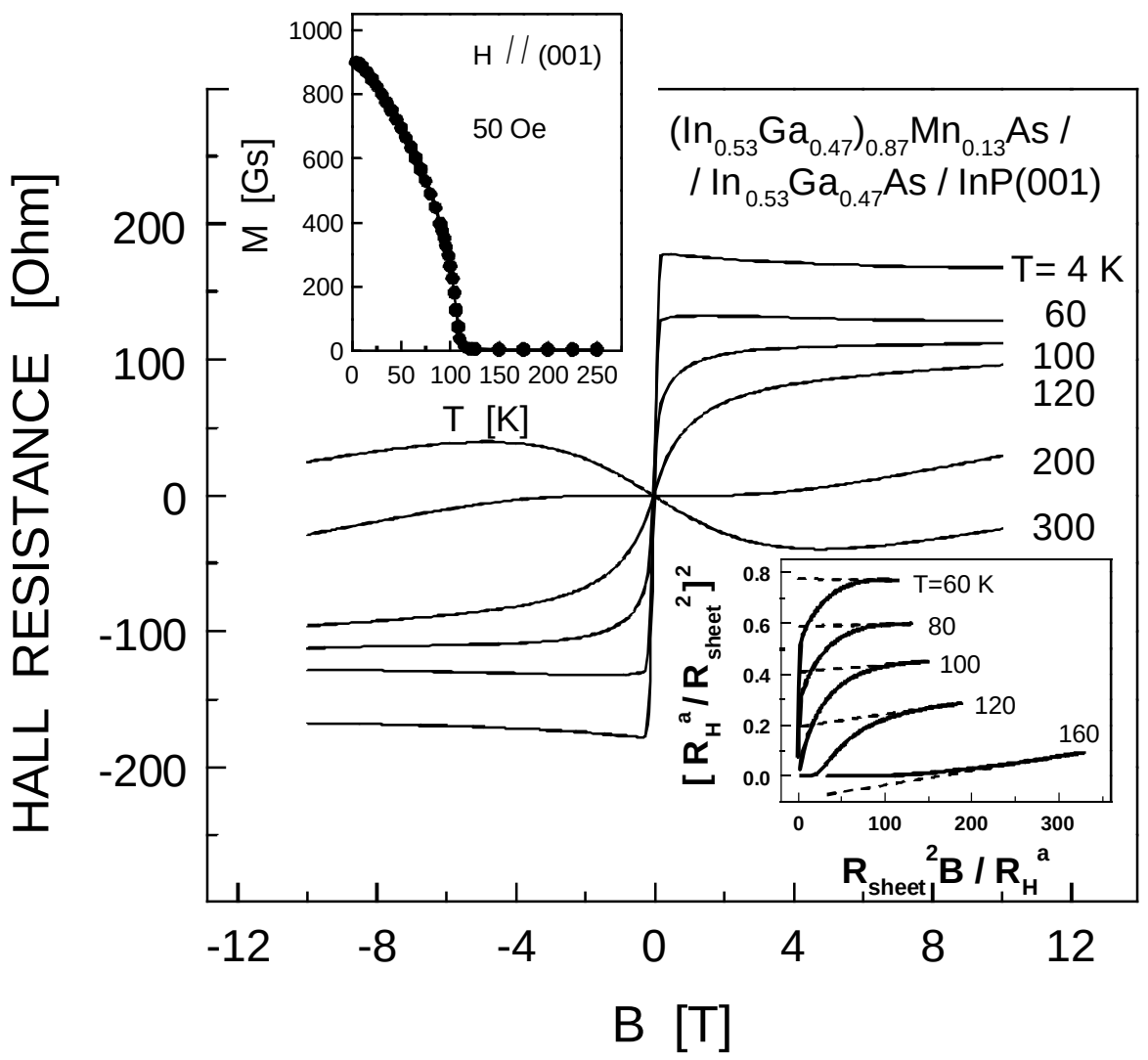

Fig.2.

T. Slupinski et al., APL

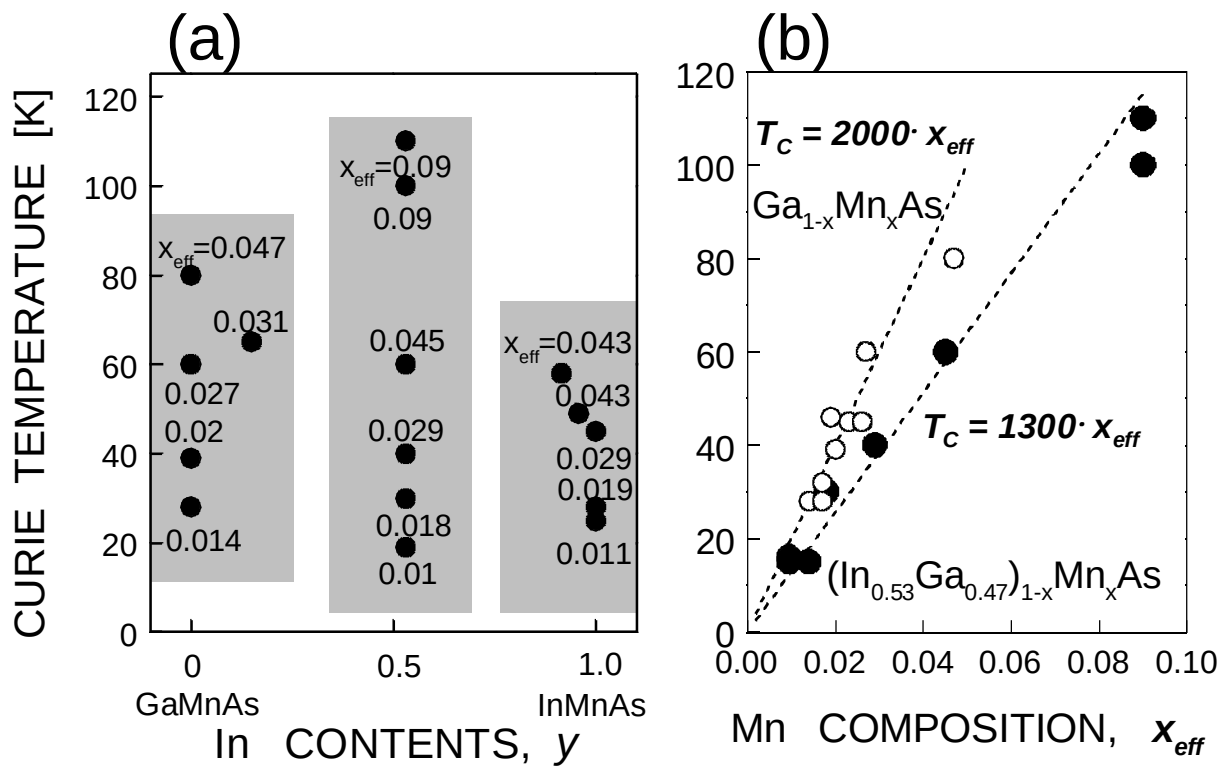

Fig.3.

T. Slupinski et al., APL

**References**

[1] H. Munekata, H. Ohno, S. von Molnár, A. Segmüller, L. L. Chang, and L. Esaki, Phys. Rev. Lett. **63**, 1849 (1989); H. Ohno, H. Munekata, T. Penney, S. von Molnár, and L. L. Chang, Phys. Rev. Lett. **68**, 2664 (1992).

[2] S. Koshihara, A. Oiwa, M. Hirasawa, S. Katsumoto, Y. Iye, C. Urano, H. Takagi and H. Munekata, Phys. Rev. Lett. **78**, 4617 (1997).

[3] H. Ohno, D. Chiba, F. Matsukura, T. Omiya, E. Abe, T. Dietl, Y. Ohno, and K. Ohtani, Nature **408**, 944 (2000).

[4] H. Ohno, J. Magn. Magn. Mater. **200**, 110 (1999); H. Ohno and F. Matsukura, Solid State Commun. **117**, 179 (2001); and references therein.

[5] H. Ohno, Science **281**, 951 (1998); T. Dietl, H. Ohno, F. Matsukura, Phys. Rev. B **63**, 195205 (2001); and references therein.

[6] H. Munekata, A. Zaslavsky, P. Fumagali and R. J. Gambino, Appl. Phys. Lett. **63**, 2929 (1993).

[7] T. Hayashi, Y. Hashimoto, S. Katsumoto and Y. Iye, Appl. Phys. Lett. **78**, 1691 (2001).

[8] K. Y. Cheng, A. Y. Cho, W. R. Wagner, and W. A. Bonner, J. Appl. Phys. **52**, 1015 (1981).

[9] H. Küntzel, J. Böttcher, R. Gibis and G. Urmann, Appl. Phys. Lett. **61**, 1347 (1992).

[10] T. Slupinski, A. Oiwa, S. Yanagi and H. Munekata, J. Cryst. Growth (2002), in press; T. Slupinski, A. Oiwa and H. Munekata, J. Cryst. Growth (2002), in pr ess.

[11] A. Oiwa, A. Endo, S. Katsumoto, Y. Iye, H. Ohno and H. Munekata, Phys. Rev. B **59**, 5826 (1999).

[12] we used $M_r$(//) measured with an in-plane magnetic field, which is very close to the low-field saturation values of both in-plane and perpendicular magnetizations.

[13] D. J. Chadi, P. H. Citrin, C. H. Park, D. L. Adler, M. A. Marcus and H.-J. Gossmann, Phys. Rev. Lett. **79**, 4834 (1997).

[14] G. Glass, H. Kim, P. Desjardins, N. Taylor, T. Spila, Q. Lu and J. E. Greene, Phys. Rev. B **61**, 7628 (2000).

[15] J. Wagner, R. C. Newman, B. R. Davidson, S. P. Westwater, T. J. Bullough, T. B. Joyce, C. D. Latham, R. Jones and S. Öberg, Phys. Rev. Lett. **78**, 74 (1997).

[16] T. Slupinski and E. Zielinska-Rohozinska, Materials Research Society Symposium Proceedings vol. 583, page 261, (2000), edited by A. Mascarenhas, D. Follstaedt, T. Suzuki, B. Joyce.

[17] $x_{eff}$ is determined from the saturation value of the ferromagnetic part of the magnetization measured at small fields $B$ ~ 0.1 Tesla along an easy axis, assuming Mn spin 5/2. Direction of $B$ is parallel to the sample plane for (In,Ga,Mn)As and (Ga,Mn)As, whereas it is perpendicular for (In,Mn)As).

[18] R. K. Kawakami, E. Johnston-Halperin, L. F. Chen, M. Hanson, N. Guebels, J. S. Speck, A. C. Gossard and D. D. Awschalom, Appl. Phys. Lett. **77**, 2379 (2000).